\documentclass[english]{iopart}
\usepackage[T1]{fontenc}
\usepackage[latin1]{inputenc}
\usepackage{babel}
\usepackage{amsfonts}
\usepackage{amssymb}
\usepackage{graphics}
\usepackage{graphicx}
\usepackage[T1]{fontenc}
\usepackage[latin1]{inputenc}
\usepackage{epsf}
\usepackage{color}

\newdimen\figwidth
\figwidth=\textwidth

{\count255=\time\divide\count255 by 60 \xdef\hourmin{\number\count255}
        \multiply\count255 by-60\advance\count255 by\time
   \xdef\hourmin{\hourmin:\ifnum\count255<10 0\fi\the\count255}}

\def\oramin{\hourmin }

\def\ladate{\number\day/\ifcase\month\or janvier \or fevrier \or mars \or
avril \or mai \or juin \or juillet \or ao\^ut \or septembre
\or octobre \or novembre \or d\'ecembre \fi/\number\year\ (\oramin)}

\setbox200\hbox{$\scriptstyle \ladate $}

\begin{document}

\title{Algebraic entropy for lattice equations}
\author{C.-M.~Viallet } \address{ Laboratoire de Physique
Th\'eorique et des Hautes Energies\\ UMR 7589 Centre National de la
Recherche Scientifique \\ Bo\^\i te 126, 4 Place Jussieu, 75252
Paris Cedex 05, France} \ead{viallet@lpthe.jussieu.fr}


\begin{abstract}
We give the basic definition of algebraic entropy for lattice
equations. The entropy is a canonical measure of the complexity of the
dynamics they define. Its vanishing is a signal of integrability, and
can be used as a powerful integrability detector. It is also
conjectured to take remarkable values (algebraic integers).
\end{abstract}

The analysis of discrete dynamical systems, in particular the measure
of their complexity, and possibly the detection of their integrability
is a huge subject\footnote[1]{We will not dwell here upon the
definition of integrability}, originating in the work of Poincar\'e.
It contains the study of the dynamics of rational maps, already a vast
topic of research. It also contains the study of lattice equations,
which are to maps what partial differential equations are to ordinary
differential equations.  Our purpose here is to extend the notion of
algebraic entropy, already widely used for
maps~\cite{FaVi93,BeVi99}~\cite{Ar90,Ve92}, and recognized as an
unmatched integrability detector~\cite{HiVi98}, to lattice equations as in~\cite{TrGrRa01},
thus introducing a measure of complexity for higher dimensional
discrete dynamics.
\bigskip

We briefly describe the setting, the space of initial data, the
evolutions, and define the related entropies. We give examples, some
integrable, some not integrable, showing how to extract information
about the global (and asymptotic) behaviour of the system from a few
iterates. We formalize and confirm the results of~\cite{TrGrRa01}.
We also present some conjectures and perspectives.

\section{The setting}
Consider a cubic lattice of dimension $D$. The vertices of the lattice
 are labeled by $D$ relative integers $[n_1, n_2,\dots, n_D]$. To each
 vertex is associated the variable $y_{[n_1, n_2, \dots, n_D]}$.  We
 are given a defining relation, which links the values of $y$ at each
 corner of the elementary cells (square for $D=2$, cube for $D=3$ and
 so on).

We will suppose that the defining relations allow to calculate any
corner values on a cell from the $2^D-1$ remaining ones, and that the
value is given by a rational expression. This implies that our
defining relations are multilinear, which covers a large number of
interesting cases (see for example~\cite{Hi81,Mi82} and more in the
last sections). This restriction may be partially lifted.

As an illustration, consider the two-dimensional case of the square
plane lattice. The elementary cell is a plaquette shown in Figure
({{1}}).

\vskip 1cm
\setlength{\unitlength}{2500sp}%
\begingroup\makeatletter\ifx\SetFigFont\undefined%
\gdef\SetFigFont#1#2#3#4#5{%
  \reset@font\fontsize{#1}{#2pt}%
  \fontfamily{#3}\fontseries{#4}\fontshape{#5}%
  \selectfont}%
\fi\endgroup%
\begin{center}
\begin{picture}(1827,1741)(211,-1385)
\label{plaquette}
{\color[rgb]{0,0,0}\thinlines
\put(901,-61){\circle{202}}
}%
{\color[rgb]{0,0,0}\put(1801,-61){\circle{202}}
}%
{\color[rgb]{0,0,0}\put(1801,-961){\circle{202}}
}%
{\color[rgb]{0,0,0}\put(901,-961){\circle{202}}
}%
{\color[rgb]{0,0,0}\put(901,164){\line( 0,-1){1395}}
}%
{\color[rgb]{0,0,0}\put(1801,164){\line( 0,-1){1395}}
}%
{\color[rgb]{0,0,0}\put(631,-61){\line( 1, 0){1395}}
}%
{\color[rgb]{0,0,0}\put(631,-961){\line( 1, 0){1395}}
}%
\put(226,209){\makebox(0,0)[lb]{\smash{{\SetFigFont{10}{14.4}{\rmdefault}{\mddefault}{\updefault}{\color[rgb]{0,0,0}$[n_1,n_2+1]$}%
}}}}
\put(406,-1321){\makebox(0,0)[lb]{\smash{{\SetFigFont{10}{14.4}{\rmdefault}{\mddefault}{\updefault}{\color[rgb]{0,0,0}$[n_1,n_2]$}%
}}}}
\put(1936,-1276){\makebox(0,0)[lb]{\smash{{\SetFigFont{10}{14.4}{\rmdefault}{\mddefault}{\updefault}{\color[rgb]{0,0,0}$[n_1+1,n_2]$}%
}}}}
\put(1891,209){\makebox(0,0)[lb]{\smash{{\SetFigFont{10}{14.4}{\rmdefault}{\mddefault}{\updefault}{\color[rgb]{0,0,0}$[n_1+1,n_2+1]$}%
}}}}
\end{picture}%
\par\smallskip
Figure {{1}}: elementary cell in two dimensions
\end{center}

The  defining relation is a constraint of the form
\begin{eqnarray}
f(y_{[n_1,n_2]},y_{[n_1+1,n_2]}, y_{[n_1,n_2+1]},y_{[n_1+1,n_2+1]}) = 0.
\end{eqnarray}

We will use some specific examples later.

\section{Initial conditions}

For the sake of clarity, we will concentrate on the $D=2$ case, but
everything we will say generalizes straightforwardly to higher
dimensions.

In order to define an evolution we have to specify initial
conditions. From the form of the defining relation, it appears that
the values of $y$ have to be given on some ``diagonal'' of the
lattice.
 
The space of initial conditions is infinite dimensional.

When $D=2$, the diagonals need to go from $[n_1=-\infty, n_2=-\infty]$
to $[n_1=\infty, n_2=\infty]$, or from $[n_1=-\infty, n_2=+\infty]$ to
$[n_1=\infty, n_2=-\infty]$.  We will restrict ourselves to regular
diagonals which are staircases with steps of {\em constant} horizontal
length, and {\em constant} height.  Figure ({{2}}) shows four
diagonals. The ones labeled $(1)$ and $(2)$ are regular. The one
labeled $(3)$ would be acceptable, but we will not consider such
diagonals. Line $(4)$ is excluded since it may lead to
incompatibilities.

\bigskip
\setlength{\unitlength}{1200sp} 
\begingroup\makeatletter\ifx\SetFigFont\undefined%
\gdef\SetFigFont#1#2#3#4#5{%
  \reset@font\fontsize{#1}{#2pt}%
  \fontfamily{#3}\fontseries{#4}\fontshape{#5}%
  \selectfont}%
\fi\endgroup%
\begin{center}
\begin{picture}(12708,9558)(397,-9115)
\label{possible}
\thinlines
{\color[rgb]{.8,.8,.8}\put(901,389){\line( 0,-1){9450}}
}%
{\color[rgb]{.8,.8,.8}\put(1351,389){\line( 0,-1){9450}}
}%
{\color[rgb]{.8,.8,.8}\put(1801,389){\line( 0,-1){9450}}
}%
{\color[rgb]{.8,.8,.8}\put(2251,389){\line( 0,-1){9450}}
}%
{\color[rgb]{.8,.8,.8}\put(2701,389){\line( 0,-1){9450}}
}%
{\color[rgb]{.8,.8,.8}\put(3151,389){\line( 0,-1){9450}}
}%
{\color[rgb]{.8,.8,.8}\put(3601,389){\line( 0,-1){9450}}
}%
{\color[rgb]{.8,.8,.8}\put(4051,389){\line( 0,-1){9450}}
}%
{\color[rgb]{.8,.8,.8}\put(4501,389){\line( 0,-1){9450}}
}%
{\color[rgb]{.8,.8,.8}\put(4951,389){\line( 0,-1){9450}}
}%
{\color[rgb]{.8,.8,.8}\put(5401,389){\line( 0,-1){9450}}
}%
{\color[rgb]{.8,.8,.8}\put(5851,389){\line( 0,-1){9450}}
}%
{\color[rgb]{.8,.8,.8}\put(6301,389){\line( 0,-1){9450}}
}%
{\color[rgb]{.8,.8,.8}\put(6751,389){\line( 0,-1){9450}}
}%
{\color[rgb]{.8,.8,.8}\put(7201,389){\line( 0,-1){9450}}
}%
{\color[rgb]{.8,.8,.8}\put(7651,389){\line( 0,-1){9450}}
}%
{\color[rgb]{.8,.8,.8}\put(8101,389){\line( 0,-1){9450}}
}%
{\color[rgb]{.8,.8,.8}\put(8551,389){\line( 0,-1){9450}}
}%
{\color[rgb]{.8,.8,.8}\put(9001,389){\line( 0,-1){9450}}
}%
{\color[rgb]{.8,.8,.8}\put(9451,389){\line( 0,-1){9450}}
}%
{\color[rgb]{.8,.8,.8}\put(9901,389){\line( 0,-1){9450}}
}%
{\color[rgb]{.8,.8,.8}\put(10351,389){\line( 0,-1){9450}}
}%
{\color[rgb]{.8,.8,.8}\put(10801,389){\line( 0,-1){9450}}
}%
{\color[rgb]{.8,.8,.8}\put(11251,389){\line( 0,-1){9450}}
}%
{\color[rgb]{.8,.8,.8}\put(11701,389){\line( 0,-1){9450}}
}%
{\color[rgb]{.8,.8,.8}\put(12151,389){\line( 0,-1){9450}}
}%
{\color[rgb]{.8,.8,.8}\put(12601,389){\line( 0,-1){9450}}
}%
{\color[rgb]{.8,.8,.8}\put(451,-61){\line( 1, 0){12600}}
}%
{\color[rgb]{.8,.8,.8}\put(451,-61){\line( 1, 0){12600}}
}%
{\color[rgb]{.8,.8,.8}\put(451,-61){\line( 1, 0){12600}}
}%
{\color[rgb]{.8,.8,.8}\put(451,-61){\line( 1, 0){12600}}
}%
{\color[rgb]{.8,.8,.8}\put(451,-511){\line( 1, 0){12600}}
}%
{\color[rgb]{.8,.8,.8}\put(451,-961){\line( 1, 0){12600}}
}%
{\color[rgb]{.8,.8,.8}\put(451,-1411){\line( 1, 0){12600}}
}%
{\color[rgb]{.8,.8,.8}\put(451,-1861){\line( 1, 0){12600}}
}%
{\color[rgb]{.8,.8,.8}\put(451,-2311){\line( 1, 0){12600}}
}%
{\color[rgb]{.8,.8,.8}\put(451,-2761){\line( 1, 0){12600}}
}%
{\color[rgb]{.8,.8,.8}\put(451,-3211){\line( 1, 0){12600}}
}%
{\color[rgb]{.8,.8,.8}\put(451,-3661){\line( 1, 0){12600}}
}%
{\color[rgb]{.8,.8,.8}\put(451,-4111){\line( 1, 0){12600}}
}%
{\color[rgb]{.8,.8,.8}\put(451,-4561){\line( 1, 0){12600}}
}%
{\color[rgb]{.8,.8,.8}\put(451,-5011){\line( 1, 0){12600}}
}%
{\color[rgb]{.8,.8,.8}\put(451,-5461){\line( 1, 0){12600}}
}%
{\color[rgb]{.8,.8,.8}\put(451,-5911){\line( 1, 0){12600}}
}%
{\color[rgb]{.8,.8,.8}\put(451,-6361){\line( 1, 0){12600}}
}%
{\color[rgb]{.8,.8,.8}\put(451,-6811){\line( 1, 0){12600}}
}%
{\color[rgb]{.8,.8,.8}\put(451,-7261){\line( 1, 0){12600}}
}%
{\color[rgb]{.8,.8,.8}\put(451,-7711){\line( 1, 0){12600}}
}%
{\color[rgb]{.8,.8,.8}\put(451,-8161){\line( 1, 0){12600}}
}%
{\color[rgb]{.8,.8,.8}\put(451,-8611){\line( 1, 0){12600}}
}%
\thicklines
{\color[rgb]{1,0,0}\put(6301,389){\line( 0,-1){450}}
\put(6301,-61){\line( 1, 0){450}}
\put(6751,-61){\line( 0,-1){450}}
\put(6751,-511){\line( 1, 0){450}}
\put(7201,-511){\line( 0,-1){450}}
\put(7201,-961){\line( 1, 0){450}}
\put(7651,-961){\line( 0,-1){450}}
\put(7651,-1411){\line( 1, 0){450}}
\put(8101,-1411){\line( 0,-1){450}}
\put(8101,-1861){\line( 1, 0){450}}
\put(8551,-1861){\line( 0,-1){450}}
\put(8551,-2311){\line( 1, 0){450}}
\put(9001,-2311){\line( 0,-1){450}}
\put(9001,-2761){\line( 1, 0){450}}
\put(9451,-2761){\line( 0,-1){450}}
\put(9451,-3211){\line( 1, 0){450}}
\put(9901,-3211){\line( 0,-1){450}}
\put(9901,-3661){\line( 1, 0){450}}
\put(10351,-3661){\line( 0,-1){450}}
\put(10351,-4111){\line( 1, 0){450}}
\put(10801,-4111){\line( 0,-1){450}}
\put(10801,-4561){\line( 1, 0){450}}
\put(11251,-4561){\line( 0,-1){450}}
\put(11251,-5011){\line( 1, 0){450}}
\put(11701,-5011){\line( 0,-1){450}}
\put(11701,-5461){\line( 1, 0){450}}
\put(12151,-5461){\line( 0,-1){450}}
\put(12151,-5911){\line( 1, 0){450}}
\put(12601,-5911){\line( 0,-1){450}}
\put(12601,-6361){\line( 1, 0){450}}
}%
\put(1000,-500){\makebox(0,0)[lb]{\smash{{\SetFigFont{20}{24.0}{\rmdefault}{\mddefault}{\updefault}{\color[rgb]{1,0,0}$(1)$}}}}}
\put(7200,-500){\makebox(0,0)[lb]{\smash{{\SetFigFont{20}{24.0}{\rmdefault}{\mddefault}{\updefault}{\color[rgb]{1,0,0}$(2)$}}}}}
\put(3600,-3300){\makebox(0,0)[lb]{\smash{{\SetFigFont{20}{24.0}{\rmdefault}{\mddefault}{\updefault}{\color[rgb]{0,0,1}$(3)$}}}}}
\put(7300,-5500){\makebox(0,0)[lb]{\smash{{\SetFigFont{20}{24.0}{\rmdefault}{\mddefault}{\updefault}{\color[rgb]{0,0,1}$(4)$}}}}}

{\color[rgb]{1,0,0}\put(901,389){\line( 0,-1){1350}}
\put(901,-961){\line( 1, 0){450}}
\put(1351,-961){\line( 0,-1){1350}}
\put(1351,-2311){\line( 1, 0){450}}
\put(1801,-2311){\line( 0,-1){1350}}
\put(1801,-3661){\line( 1, 0){450}}
\put(2251,-3661){\line( 0,-1){1350}}
\put(2251,-5011){\line( 1, 0){450}}
\put(2701,-5011){\line( 0,-1){1350}}
\put(2701,-6361){\line( 1, 0){450}}
\put(3151,-6361){\line( 0,-1){1350}}
\put(3151,-7711){\line( 1, 0){450}}
\put(3601,-7711){\line( 0,-1){1350}}
}%
{\color[rgb]{0,0,1}\put(451,-4561){\line( 1, 0){900}}
\put(1351,-4561){\line( 0, 1){450}}
\put(1351,-4111){\line( 1, 0){1800}}
\put(3151,-4111){\line( 0, 1){1800}}
\put(3151,-2311){\line( 1, 0){2250}}
\put(5401,-2311){\line( 0, 1){450}}
\put(5401,-1861){\line( 1, 0){4050}}
\put(9451,-1861){\line( 0, 1){450}}
\put(9451,-1411){\line( 1, 0){450}}
\put(9901,-1411){\line( 0, 1){450}}
\put(9901,-961){\line( 1, 0){450}}
\put(10351,-961){\line( 0, 1){900}}
\put(10351,-61){\line( 1, 0){900}}
\put(11251,-61){\line( 0, 1){450}}
}%
{\color[rgb]{0,0,1}\put(451,-8611){\line( 1, 0){450}}
\put(901,-8611){\line( 0, 1){900}}
\put(901,-7711){\line( 1, 0){450}}
\put(1351,-7711){\line( 0, 1){900}}
\put(1351,-6811){\line( 1, 0){900}}
\put(2251,-6811){\line( 0,-1){900}}
\put(2251,-7711){\line( 1, 0){450}}
\put(2701,-7711){\line( 0, 1){900}}
\put(2701,-6811){\line( 1, 0){900}}
\put(3601,-6811){\line( 0, 1){1350}}
\put(3601,-5461){\line( 1, 0){1350}}
\put(4951,-5461){\line( 0,-1){450}}
\put(4951,-5911){\line( 1, 0){1350}}
\put(6301,-5911){\line( 0, 1){900}}
\put(6301,-5011){\line( 1, 0){900}}
\put(7201,-5011){\line( 0, 1){900}}
\put(7201,-4111){\line( 1, 0){900}}
\put(8101,-4111){\line( 0, 1){1350}}
\put(8101,-2761){\line( 1, 0){2250}}
\put(10351,-2761){\line( 0,-1){4950}}
\put(10351,-7711){\line( 1, 0){1350}}
\put(11701,-7711){\line( 0, 1){6300}}
\put(11701,-1411){\line( 1, 0){450}}
\put(12151,-1411){\line( 0, 1){450}}
\put(12151,-961){\line( 1, 0){900}}
}%
\end{picture}%
\par\smallskip
Figure {{2}}: diagonals
\end{center}

\section{Restricted initial conditions}

Here again we use the $D=2$ example, to make things simple.  Given a
line of initial conditions, it is possible to calculate the values $y$
all over the $D$-dimensional space.  We have a well defined evolution,
since we restrict ourselves to regular diagonals.  Moreover, and this
is a crucial point, {\em if we want to evaluate the transformation
formula for a finite number of iterations, we only need a diagonal of
initial conditions with finite extent}.

For any positive integer $N$, and each pair of relative integers
${[\lambda_1,\lambda_2]}$, we denote by
$\Delta_{[\lambda_1,\lambda_2]}^{(N)}$, a regular diagonal consisting
of $N$ steps, each having horizontal size $l_1 = \vert \lambda_1
\vert$, height $l_2 = \vert \lambda_2 \vert$, and going in the
direction of positive (resp. negative) $n_k$, if $\lambda_k > 0$
(resp.  $\lambda_k < 0$), for $k=1..D=2$. See Figure
({{3}}).

\bigskip
\begin{center}
\setlength{\unitlength}{1700sp}%
\begingroup\makeatletter\ifx\SetFigFont\undefined%
\gdef\SetFigFont#1#2#3#4#5{%
  \reset@font\fontsize{#1}{#2pt}%
  \fontfamily{#3}\fontseries{#4}\fontshape{#5}%
  \selectfont}%
\fi\endgroup%
\begin{picture}(8536,4934)(1733,-4628)
\label{restricted}
{\color[rgb]{0,0,0}\thinlines
\put(3001,239){\circle{120}}
}%
{\color[rgb]{0,0,0}\put(3001,-961){\circle{120}}
}%
{\color[rgb]{0,0,0}\put(3601,-961){\circle{120}}
}%
{\color[rgb]{0,0,0}\put(3601,-2161){\circle{120}}
}%
{\color[rgb]{0,0,0}\put(4201,-2161){\circle{120}}
}%
{\color[rgb]{0,0,0}\put(4201,-3361){\circle{120}}
}%
{\color[rgb]{0,0,0}\put(4801,-3361){\circle{120}}
}%
{\color[rgb]{0,0,0}\put(4801,-4561){\circle{120}}
}%
{\color[rgb]{0,0,0}\put(5401,-4561){\circle{120}}
}%
{\color[rgb]{0,0,0}\put(9601,-961){\circle{120}}
}%
{\color[rgb]{0,0,0}\put(6601,-4561){\circle{120}}
}%
{\color[rgb]{0,0,0}\put(6601,-3961){\circle{120}}
}%
{\color[rgb]{0,0,0}\put(8401,-3961){\circle{120}}
}%
{\color[rgb]{0,0,0}\put(8401,-3361){\circle{120}}
}%
{\color[rgb]{0,0,0}\put(10201,-3361){\circle{120}}
}%
{\color[rgb]{0,0,0}\put(3001,-3961){\circle{120}}
}%
{\color[rgb]{0,0,0}\put(3001,-4561){\circle{120}}
}%
{\color[rgb]{0,0,0}\put(3601,-4561){\circle{120}}
}%
{\color[rgb]{0,0,0}\put(2401,-3961){\circle{120}}
}%
{\color[rgb]{0,0,0}\put(2401,-3361){\circle{120}}
}%
{\color[rgb]{0,0,0}\put(1801,-3361){\circle{120}}
}%
{\color[rgb]{0,0,0}\put(1801,-2761){\circle{120}}
}%
{\color[rgb]{0,0,0}\put(5401,239){\circle{120}}
}%
{\color[rgb]{0,0,0}\put(6001,239){\circle{120}}
}%
{\color[rgb]{0,0,0}\put(6001,-361){\circle{120}}
}%
{\color[rgb]{0,0,0}\put(6601,-361){\circle{120}}
}%
{\color[rgb]{0,0,0}\put(6601,-961){\circle{120}}
}%
{\color[rgb]{0,0,0}\put(7201,-961){\circle{120}}
}%
{\color[rgb]{0,0,0}\put(7201,-1561){\circle{120}}
}%
{\color[rgb]{0,0,0}\put(9001,-961){\circle{120}}
}%
{\color[rgb]{0,0,0}\put(9601,-361){\circle{120}}
}%
{\color[rgb]{0,0,0}\put(10201,-361){\circle{120}}
}%
{\color[rgb]{0,0,0}\put(10201,239){\circle{120}}
}%
{\color[rgb]{0,0,0}\put(7201,-3961){\circle{120}}
}%
{\color[rgb]{0,0,0}\put(7801,-3961){\circle{120}}
}%
{\color[rgb]{0,0,0}\put(9001,-3361){\circle{120}}
}%
{\color[rgb]{0,0,0}\put(9601,-3361){\circle{120}}
}%
{\color[rgb]{0,0,0}\put(4801,-3961){\circle{120}}
}%
{\color[rgb]{0,0,0}\put(4201,-2761){\circle{120}}
}%
{\color[rgb]{0,0,0}\put(3601,-1561){\circle{120}}
}%
{\color[rgb]{0,0,0}\put(3001,-361){\circle{120}}
}%
{\color[rgb]{0,0,0}\put(3001,239){\line( 0,-1){1200}}
}%
{\color[rgb]{0,0,0}\put(3001,239){\line( 0,-1){1200}}
\put(3001,-961){\line( 1, 0){600}}
\put(3601,-961){\line( 0,-1){1200}}
\put(3601,-2161){\line( 1, 0){600}}
\put(4201,-2161){\line( 0,-1){1200}}
\put(4201,-3361){\line( 1, 0){600}}
\put(4801,-3361){\line( 0,-1){1200}}
\put(4801,-4561){\line( 1, 0){600}}
}%
{\color[rgb]{0,0,0}\put(6601,-4561){\line( 0, 1){600}}
\put(6601,-3961){\line( 1, 0){1800}}
\put(8401,-3961){\line( 0, 1){600}}
\put(8401,-3361){\line( 1, 0){1800}}
}%
{\color[rgb]{0,0,0}\put(1801,-2761){\line( 0,-1){600}}
\put(1801,-3361){\line( 1, 0){600}}
\put(2401,-3361){\line( 0,-1){600}}
\put(2401,-3961){\line( 1, 0){600}}
\put(3001,-3961){\line( 0,-1){600}}
\put(3001,-4561){\line( 1, 0){600}}
}%
{\color[rgb]{0,0,0}\put(5401,239){\line( 1, 0){600}}
\put(6001,239){\line( 0,-1){600}}
\put(6001,-361){\line( 1, 0){600}}
\put(6601,-361){\line( 0,-1){600}}
\put(6601,-961){\line( 1, 0){600}}
\put(7201,-961){\line( 0,-1){600}}
}%
{\color[rgb]{0,0,0}\put(9001,-961){\line( 1, 0){600}}
\put(9601,-961){\line( 0, 1){600}}
\put(9601,-361){\line( 1, 0){600}}
\put(10201,-361){\line( 0, 1){600}}
}%
{\color[rgb]{1,0,0}\put(6601,-2644){\vector(0,1){600}}}
{\color[rgb]{1,0,0}\put(6601,-2644){\vector(1,0){600}}}
\put(400,-4336){\makebox(0,0)[lb]{\smash{{\SetFigFont{20}{24.0}{\rmdefault}{\mddefault}{\updefault}{\color[rgb]{0,0,0}$\Delta_{(-1,1)}^{(3)}$}%
}}}}
\put(3451,-361){\makebox(0,0)[lb]{\smash{{\SetFigFont{20}{24.0}{\rmdefault}{\mddefault}{\updefault}{\color[rgb]{0,0,0}$\Delta_{(-1,2)}^{(4)}$}%
}}}}
\put(6610, 14){\makebox(0,0)[lb]{\smash{{\SetFigFont{20}{24.0}{\rmdefault}{\mddefault}{\updefault}{\color[rgb]{0,0,0}$\Delta_{(1,-1)}^{(3)}$}%
}}}}
\put(9226,-1600){\makebox(0,0)[lb]{\smash{{\SetFigFont{20}{24.0}{\rmdefault}{\mddefault}{\updefault}{\color[rgb]{0,0,0}$\Delta_{(1,1)}^{(2)}$}%
}}}}
\put(8776,-4336){\makebox(0,0)[lb]{\smash{{\SetFigFont{20}{24.0}{\rmdefault}{\mddefault}{\updefault}{\color[rgb]{0,0,0}$\Delta_{(-3,-1)}^{(2)}$}%
}}}}
\end{picture}%

\par\bigskip
Figure {{3}}: various choices of restricted initial conditions
\end{center}

Suppose we fix the initial conditions on
$\Delta_{[\lambda_1,\lambda_2]}^{(N)}$. We may calculate $y$ over a
rectangle of size $(N l_1+1) \times (N l_2 +1)$. The diagonal cuts the
rectangle in two halves.  One of them uses all initial values, and we
will calculate the evolution only on that part. See
Figure({{4}}).

\setlength{\unitlength}{2400sp}%
\begingroup\makeatletter\ifx\SetFigFont\undefined%
\gdef\SetFigFont#1#2#3#4#5{%
  \reset@font\fontsize{#1}{#2pt}%
  \fontfamily{#3}\fontseries{#4}\fontshape{#5}%
  \selectfont}%
\fi\endgroup%

\bigskip

\begin{center}
\begin{picture}(1940,3738)(2933,-3428)
\label{range}
{
\color[rgb]{0,0,0}\thinlines
\put(3001,239){\circle{120}}
}%
{\color[rgb]{0,0,0}\put(3001,-961){\circle{120}}
}%
{\color[rgb]{0,0,0}\put(3601,-961){\circle{120}}
}%
{\color[rgb]{0,0,0}\put(3601,-2161){\circle{120}}
}%
{\color[rgb]{0,0,0}\put(4201,-2161){\circle{120}}
}%
{\color[rgb]{0,0,0}\put(4201,-3361){\circle{120}}
}%
{\color[rgb]{0,0,0}\put(4801,-3361){\circle{120}}
}%
{\color[rgb]{0,0,0}\put(4201,-2761){\circle{120}}
}%
{\color[rgb]{0,0,0}\put(3601,-1561){\circle{120}}
}%
{\color[rgb]{0,0,0}\put(3001,-361){\circle{120}}
}%
{\color[rgb]{0,0,0}\put(3601,239){\circle*{128}}
}%
{\color[rgb]{0,0,0}\put(3601,-361){\circle*{128}}
}%
{\color[rgb]{0,0,0}\put(4201,-361){\circle*{128}}
}%
{\color[rgb]{0,0,0}\put(4201,239){\circle*{128}}
}%
{\color[rgb]{0,0,0}\put(4801,239){\circle*{128}}
}%
{\color[rgb]{0,0,0}\put(4801,-361){\circle*{128}}
}%
{\color[rgb]{0,0,0}\put(4801,-961){\circle*{128}}
}%
{\color[rgb]{0,0,0}\put(4201,-961){\circle*{128}}
}%
{\color[rgb]{0,0,0}\put(4201,-1561){\circle*{128}}
}%
{\color[rgb]{0,0,0}\put(4801,-1561){\circle*{128}}
}%
{\color[rgb]{0,0,0}\put(4801,-2161){\circle*{128}}
}%
{\color[rgb]{0,0,0}\put(4801,-2761){\circle*{128}}
}%
{\color[rgb]{0,0,0}\put(3001,239){\line( 0,-1){1200}}
}%
{\color[rgb]{0,0,0}\put(4201,-2161){\line( 1, 0){600}}
}%
{\color[rgb]{0,0,0}\put(3601,-1561){\line( 1, 0){1200}}
}%
{\color[rgb]{0,0,0}\put(3601,-961){\line( 1, 0){1200}}
}%
{\color[rgb]{0,0,0}\put(3001,-361){\line( 1, 0){1800}}
}%
{\color[rgb]{0,0,0}\put(3601,239){\line( 0,-1){1200}}
}%
{\color[rgb]{0,0,0}\put(4201,239){\line( 0,-1){2400}}
}%
{\color[rgb]{0,0,0}\put(4801,239){\line( 0,-1){3600}}
}%
{\color[rgb]{0,0,0}\put(4201,-2761){\line( 1, 0){600}}
}%
{\color[rgb]{0,0,0}\put(3001,239){\line( 1, 0){1800}}
}%
{\color[rgb]{0,0,0}\put(3001,-961){\line( 1, 0){600}}
\put(3601,-961){\line( 0,-1){600}}
\put(3601,-1561){\line( 0,-1){600}}
\put(3601,-2161){\line( 1, 0){600}}
\put(4201,-2161){\line( 0,-1){1200}}
\put(4201,-3361){\line( 1, 0){600}}
}%
\end{picture}%
\par\bigskip
Figure {{4}}: range of initial conditions $\Delta_{[-1,2]}^{(3)}$
\end{center}

\section{Fundamental entropies}
\label{fundamental}
We are now in position to calculate ``iterates'' of the
evolution. Choose some restricted diagonal
$\Delta_{[\lambda_1,\lambda_2]}^{(N)}$. The total number of initial
points is $q = N( l_1 + l_2)+1$.

For such restricted initial data, the natural space where the
evolution acts is the projective space $P_q$ of dimension $q$. We may
calculate the iterates and fill Figure ({{4}}), considering the
$q$ initial values as inhomogeneous coordinates of $P_q$.

Evaluating the degrees of the successive iterates, we will produce
double sequences of degrees.

The simplest possible choice is to apply this construction to the
restricted diagonals $\Delta_{[\pm 1, \pm 1]}^{(N)}$, which we will
denote $\Delta_{++}^{(N)}$, $\Delta_{+-}^{(N)}$, \dots, and call them
fundamental diagonals (the upper index $(N)$ is omitted for infinite
lines).

The pattern of degrees is then of the form

\begin{eqnarray}
\matrix {
  1 & d^{(1)} & d^{(2)} & \dots   & d^{(N-1)} &  d^{(N)} \cr
  1 & 1       & d^{(1)} & d^{(2)} & \dots     & d^{(N-1)} \cr
    & 1       & 1       & d^{(1)} & d^{(2)}   & \dots  \cr
    &         & 1       &  1      & d^{(1)}   & d^{(2)}  \cr
    &         &         &  1      & 1         & d^{(1)} \cr
    &         &         &         & 1         & 1 \cr
}
\end{eqnarray}

To each choice of indices $[\pm 1, \pm 1]$ we associate a sequence of
degrees $d_{\pm \pm}^{(n)}$.

{\bf Definition:} The fundamental entropies of the lattice equation
are given by
\begin{eqnarray}
\label{algent}
\epsilon_{\pm  \pm } = \lim_{n\rightarrow\infty}{1\over{n}}\;
log(d^{(n)}_{\pm \pm}) .
\end{eqnarray}

{\bf Claim:} These entropies always exist~\cite{BeVi99}, because of
the subadditivity property of the logarithm of the degree of composed
maps.

The fundamental entropies correspond to initial data given on
diagonals with slope $+1$ or $-1$, and evolutions towards the four
corners of the lattice, as shown in Figure ({{5}}).

\bigskip
\begin{center}
\setlength{\unitlength}{1200sp}%
\begingroup\makeatletter\ifx\SetFigFont\undefined%
\gdef\SetFigFont#1#2#3#4#5{%
  \reset@font\fontsize{#1}{#2pt}%
  \fontfamily{#3}\fontseries{#4}\fontshape{#5}%
  \selectfont}%
\fi\endgroup%
\begin{picture}(12708,9558)(397,-9115)
\label{fourcorners}
\thinlines
{\color[rgb]{.8,.8,.8}\put(901,389){\line( 0,-1){9450}}
}%
{\color[rgb]{.8,.8,.8}\put(1351,389){\line( 0,-1){9450}}
}%
{\color[rgb]{.8,.8,.8}\put(1801,389){\line( 0,-1){9450}}
}%
{\color[rgb]{.8,.8,.8}\put(2251,389){\line( 0,-1){9450}}
}%
{\color[rgb]{.8,.8,.8}\put(2701,389){\line( 0,-1){9450}}
}%
{\color[rgb]{.8,.8,.8}\put(3151,389){\line( 0,-1){9450}}
}%
{\color[rgb]{.8,.8,.8}\put(3601,389){\line( 0,-1){9450}}
}%
{\color[rgb]{.8,.8,.8}\put(4051,389){\line( 0,-1){9450}}
}%
{\color[rgb]{.8,.8,.8}\put(4501,389){\line( 0,-1){9450}}
}%
{\color[rgb]{.8,.8,.8}\put(4951,389){\line( 0,-1){9450}}
}%
{\color[rgb]{.8,.8,.8}\put(5401,389){\line( 0,-1){9450}}
}%
{\color[rgb]{.8,.8,.8}\put(5851,389){\line( 0,-1){9450}}
}%
{\color[rgb]{.8,.8,.8}\put(6301,389){\line( 0,-1){9450}}
}%
{\color[rgb]{.8,.8,.8}\put(6751,389){\line( 0,-1){9450}}
}%
{\color[rgb]{.8,.8,.8}\put(7201,389){\line( 0,-1){9450}}
}%
{\color[rgb]{.8,.8,.8}\put(7651,389){\line( 0,-1){9450}}
}%
{\color[rgb]{.8,.8,.8}\put(8101,389){\line( 0,-1){9450}}
}%
{\color[rgb]{.8,.8,.8}\put(8551,389){\line( 0,-1){9450}}
}%
{\color[rgb]{.8,.8,.8}\put(9001,389){\line( 0,-1){9450}}
}%
{\color[rgb]{.8,.8,.8}\put(9451,389){\line( 0,-1){9450}}
}%
{\color[rgb]{.8,.8,.8}\put(9901,389){\line( 0,-1){9450}}
}%
{\color[rgb]{.8,.8,.8}\put(10351,389){\line( 0,-1){9450}}
}%
{\color[rgb]{.8,.8,.8}\put(10801,389){\line( 0,-1){9450}}
}%
{\color[rgb]{.8,.8,.8}\put(11251,389){\line( 0,-1){9450}}
}%
{\color[rgb]{.8,.8,.8}\put(11701,389){\line( 0,-1){9450}}
}%
{\color[rgb]{.8,.8,.8}\put(12151,389){\line( 0,-1){9450}}
}%
{\color[rgb]{.8,.8,.8}\put(12601,389){\line( 0,-1){9450}}
}%
{\color[rgb]{.8,.8,.8}\put(451,-61){\line( 1, 0){12600}}
}%
{\color[rgb]{.8,.8,.8}\put(451,-61){\line( 1, 0){12600}}
}%
{\color[rgb]{.8,.8,.8}\put(451,-61){\line( 1, 0){12600}}
}%
{\color[rgb]{.8,.8,.8}\put(451,-61){\line( 1, 0){12600}}
}%
{\color[rgb]{.8,.8,.8}\put(451,-511){\line( 1, 0){12600}}
}%
{\color[rgb]{.8,.8,.8}\put(451,-961){\line( 1, 0){12600}}
}%
{\color[rgb]{.8,.8,.8}\put(451,-1411){\line( 1, 0){12600}}
}%
{\color[rgb]{.8,.8,.8}\put(451,-1861){\line( 1, 0){12600}}
}%
{\color[rgb]{.8,.8,.8}\put(451,-2311){\line( 1, 0){12600}}
}%
{\color[rgb]{.8,.8,.8}\put(451,-2761){\line( 1, 0){12600}}
}%
{\color[rgb]{.8,.8,.8}\put(451,-3211){\line( 1, 0){12600}}
}%
{\color[rgb]{.8,.8,.8}\put(451,-3661){\line( 1, 0){12600}}
}%
{\color[rgb]{.8,.8,.8}\put(451,-4111){\line( 1, 0){12600}}
}%
{\color[rgb]{.8,.8,.8}\put(451,-4561){\line( 1, 0){12600}}
}%
{\color[rgb]{.8,.8,.8}\put(451,-5011){\line( 1, 0){12600}}
}%
{\color[rgb]{.8,.8,.8}\put(451,-5461){\line( 1, 0){12600}}
}%
{\color[rgb]{.8,.8,.8}\put(451,-5911){\line( 1, 0){12600}}
}%
{\color[rgb]{.8,.8,.8}\put(451,-6361){\line( 1, 0){12600}}
}%
{\color[rgb]{.8,.8,.8}\put(451,-6811){\line( 1, 0){12600}}
}%
{\color[rgb]{.8,.8,.8}\put(451,-7261){\line( 1, 0){12600}}
}%
{\color[rgb]{.8,.8,.8}\put(451,-7711){\line( 1, 0){12600}}
}%
{\color[rgb]{.8,.8,.8}\put(451,-8161){\line( 1, 0){12600}}
}%
{\color[rgb]{.8,.8,.8}\put(451,-8611){\line( 1, 0){12600}}
}%
\thicklines
{\color[rgb]{1,0,0}\put(6301,389){\line( 0,-1){450}}
\put(6301,-61){\line( 1, 0){450}}
\put(6751,-61){\line( 0,-1){450}}
\put(6751,-511){\line( 1, 0){450}}
\put(7201,-511){\line( 0,-1){450}}
\put(7201,-961){\line( 1, 0){450}}
\put(7651,-961){\line( 0,-1){450}}
\put(7651,-1411){\line( 1, 0){450}}
\put(8101,-1411){\line( 0,-1){450}}
\put(8101,-1861){\line( 1, 0){450}}
\put(8551,-1861){\line( 0,-1){450}}
\put(8551,-2311){\line( 1, 0){450}}
\put(9001,-2311){\line( 0,-1){450}}
\put(9001,-2761){\line( 1, 0){450}}
\put(9451,-2761){\line( 0,-1){450}}
\put(9451,-3211){\line( 1, 0){450}}
\put(9901,-3211){\line( 0,-1){450}}
\put(9901,-3661){\line( 1, 0){450}}
\put(10351,-3661){\line( 0,-1){450}}
\put(10351,-4111){\line( 1, 0){450}}
\put(10801,-4111){\line( 0,-1){450}}
\put(10801,-4561){\line( 1, 0){450}}
\put(11251,-4561){\line( 0,-1){450}}
\put(11251,-5011){\line( 1, 0){450}}
\put(11701,-5011){\line( 0,-1){450}}
\put(11701,-5461){\line( 1, 0){450}}
\put(12151,-5461){\line( 0,-1){450}}
\put(12151,-5911){\line( 1, 0){450}}
\put(12601,-5911){\line( 0,-1){450}}
\put(12601,-6361){\line( 1, 0){450}}
}%
{\color[rgb]{0,0,1}\put(3151,389){\line( 0,-1){450}}
\put(3151,-61){\line( 1, 0){450}}
\put(3601,-61){\line( 0,-1){450}}
\put(3601,-511){\line( 1, 0){450}}
\put(4051,-511){\line( 0,-1){450}}
\put(4051,-961){\line( 1, 0){450}}
\put(4501,-961){\line( 0,-1){450}}
\put(4501,-1411){\line( 1, 0){450}}
\put(4951,-1411){\line( 0,-1){450}}
\put(4951,-1861){\line( 1, 0){450}}
\put(5401,-1861){\line( 0,-1){450}}
\put(5401,-2311){\line( 1, 0){450}}
\put(5851,-2311){\line( 0,-1){450}}
\put(5851,-2761){\line( 1, 0){450}}
\put(6301,-2761){\line( 0,-1){450}}
\put(6301,-3211){\line( 1, 0){450}}
\put(6751,-3211){\line( 0,-1){450}}
\put(6751,-3661){\line( 1, 0){450}}
\put(7201,-3661){\line( 0,-1){450}}
\put(7201,-4111){\line( 1, 0){450}}
\put(7651,-4111){\line( 0,-1){450}}
\put(7651,-4561){\line( 1, 0){450}}
\put(8101,-4561){\line( 0,-1){450}}
\put(8101,-5011){\line( 1, 0){450}}
\put(8551,-5011){\line( 0,-1){450}}
\put(8551,-5461){\line( 1, 0){450}}
\put(9001,-5461){\line( 0,-1){450}}
\put(9001,-5911){\line( 1, 0){450}}
\put(9451,-5911){\line( 0,-1){450}}
\put(9451,-6361){\line( 1, 0){450}}
\put(9901,-6361){\line( 0,-1){450}}
\put(9901,-6811){\line( 1, 0){450}}
\put(10351,-6811){\line( 0,-1){450}}
\put(10351,-7261){\line( 1, 0){450}}
\put(10801,-7261){\line( 0,-1){450}}
\put(10801,-7711){\line( 1, 0){450}}
\put(11251,-7711){\line( 0,-1){450}}
\put(11251,-8161){\line( 1, 0){450}}
\put(11701,-8161){\line( 0,-1){450}}
\put(11701,-8611){\line( 1, 0){450}}
\put(12151,-8611){\line(-1, 0){450}}
\put(11701,-8611){\line( 1, 0){450}}
\put(12151,-8611){\line( 0,-1){450}}
}%
{\color[rgb]{0,1,0}\put(451,-5461){\line( 1, 0){450}}
\put(901,-5461){\line( 0, 1){450}}
\put(901,-5011){\line( 1, 0){450}}
\put(1351,-5011){\line( 0, 1){450}}
\put(1351,-4561){\line( 1, 0){450}}
\put(1801,-4561){\line( 0, 1){450}}
\put(1801,-4111){\line( 1, 0){450}}
\put(2251,-4111){\line( 0, 1){450}}
\put(2251,-3661){\line( 1, 0){450}}
\put(2701,-3661){\line( 0, 1){450}}
\put(2701,-3211){\line( 1, 0){450}}
\put(3151,-3211){\line( 0, 1){450}}
\put(3151,-2761){\line( 1, 0){450}}
\put(3601,-2761){\line( 0, 1){450}}
\put(3601,-2311){\line( 1, 0){450}}
\put(4051,-2311){\line( 0, 1){450}}
\put(4051,-1861){\line( 1, 0){450}}
\put(4501,-1861){\line( 0, 1){450}}
\put(4501,-1411){\line( 1, 0){450}}
\put(4951,-1411){\line( 0, 1){450}}
\put(4951,-961){\line( 1, 0){450}}
\put(5401,-961){\line( 0, 1){450}}
\put(5401,-511){\line( 1, 0){450}}
\put(5851,-511){\line( 0, 1){450}}
\put(5851,-61){\line( 1, 0){450}}
\put(6301,-61){\line( 0, 1){450}}
}%
{\color[rgb]{1,0,1}\put(12601,389){\line( 0,-1){450}}
\put(12601,-61){\line(-1, 0){450}}
\put(12151,-61){\line( 0,-1){450}}
\put(12151,-511){\line(-1, 0){450}}
\put(11701,-511){\line( 0,-1){450}}
\put(11701,-961){\line(-1, 0){450}}
\put(11251,-961){\line( 0,-1){450}}
\put(11251,-1411){\line(-1, 0){450}}
\put(10801,-1411){\line( 0,-1){450}}
\put(10801,-1861){\line(-1, 0){450}}
\put(10351,-1861){\line( 0,-1){450}}
\put(10351,-2311){\line(-1, 0){450}}
\put(9901,-2311){\line( 0,-1){450}}
\put(9901,-2761){\line(-1, 0){450}}
\put(9451,-2761){\line( 0,-1){450}}
\put(9451,-3211){\line(-1, 0){450}}
\put(9001,-3211){\line( 0,-1){450}}
\put(9001,-3661){\line(-1, 0){450}}
\put(8551,-3661){\line( 0,-1){450}}
\put(8551,-4111){\line(-1, 0){450}}
\put(8101,-4111){\line( 0,-1){450}}
\put(8101,-4561){\line(-1, 0){450}}
\put(7651,-4561){\line( 0,-1){450}}
\put(7651,-5011){\line(-1, 0){450}}
\put(7201,-5011){\line( 0,-1){450}}
\put(7201,-5461){\line(-1, 0){450}}
\put(6751,-5461){\line( 0,-1){450}}
\put(6751,-5911){\line(-1, 0){450}}
\put(6301,-5911){\line( 0,-1){450}}
\put(6301,-6361){\line(-1, 0){450}}
\put(5851,-6361){\line( 0,-1){450}}
\put(5851,-6811){\line(-1, 0){450}}
\put(5401,-6811){\line( 0,-1){450}}
\put(5401,-7261){\line(-1, 0){450}}
\put(4951,-7261){\line( 0,-1){450}}
\put(4951,-7711){\line(-1, 0){450}}
\put(4501,-7711){\line( 0,-1){450}}
\put(4501,-8161){\line(-1, 0){450}}
\put(4051,-8161){\line( 0,-1){450}}
\put(4051,-8611){\line(-1, 0){450}}
\put(3601,-8611){\line( 0,-1){450}}
}%
{\color[rgb]{1,0,1}\put(5581,-6541){\line(-1, 1){630}}
}%
{\color[rgb]{1,0,1}\put(5581,-6541){\vector(-1, 1){630}}
}%
{\color[rgb]{0,1,0}\put(1981,-4291){\vector( 1,-1){720}}
}%
{\color[rgb]{1,0,0}\put(7471,-691){\vector( 1, 1){630}}
}%
{\color[rgb]{0,0,1}\put(10531,-7531){\vector(-1,-1){630}}
}%
\put(3800,-6811){\makebox(0,0)[lb]{\smash{{\SetFigFont{25}{30.0}{\rmdefault}{\mddefault}{\updefault}{\color[rgb]{0,0,0}$\Delta_{++}$}%
}}}}
\put(2400,-4561){\makebox(0,0)[lb]{\smash{{\SetFigFont{25}{30.0}{\rmdefault}{\mddefault}{\updefault}{\color[rgb]{0,0,0}$\Delta_{--}$}%
}}}}
\put(8101,-961){\makebox(0,0)[lb]{\smash{{\SetFigFont{25}{30.0}{\rmdefault}{\mddefault}{\updefault}{\color[rgb]{0,0,0}$\Delta_{-+}$}%
}}}}
\put(8500,-7711){\makebox(0,0)[lb]{\smash{{\SetFigFont{25}{30.0}{\rmdefault}{\mddefault}{\updefault}{\color[rgb]{0,0,0}$\Delta_{+-}$}%
}}}}
\end{picture}%

\par\smallskip
Figure {{5}}: fundamental evolutions on a square lattice
\end{center}

These four entropies do not have to be identical (see section
(\ref{anisotropic})).

When the entropy vanishes, the growth of the degree is polynomial, and
the degree of that polynomial is a secondary characterization of the
complexity.

\section{Subsidiary  entropies}

We may also define entropies for the other regular diagonals. They
correspond to initial data given on a line with a slope different from
$\pm 1 $. They are useful in view of the various finite dimensional
reductions presented for example in~\cite{QuCaPaNi91}.

As an example, the pattern of degrees for $\Delta^{(3)}_{[-1,2]}$ 
looks like
\begin{eqnarray}
\left[ 
\begin {array}{cccc}
1 & d^{(1)}[1] & d^{(2)}[1] & d^{(3)}[1] = d^{(6)}[2]   \\
1 & \dots & \dots & d^{(5)}[2] \\
1 & 1 & \dots & d^{(4)}[2] \\
 & 1 & \dots & d^{(3)}[2] \\
 & 1 & 1 & d^{(2)}[2] \\
 &  & 1 & d^{(1)}[2] \\
 &  & 1 & 1 \\
\end {array} \right] 
\end{eqnarray}

The sequences of degrees we will retain are the border sequences
$\{d^{(n)}[\nu]$\}, $\nu = 1,2$, seen on the edges of the domain.
There are as many such border sequences as there are dimensions in the
lattice (here $D$=2).  The index in bracket refers to the direction of
the edge considered.  This leads to subsidiary entropies $\epsilon_{[
\lambda_1,\dots, \lambda_D ]}[\nu]$, $\nu = 1\dots D$:

\begin{eqnarray}
\label{algent2}
\epsilon_{[\lambda_1,\dots, \lambda_D] }[\nu] =
\lim_{n\rightarrow\infty}{1\over{n}}\;
\log(d^{(n)}_{[\lambda_1,\dots, \lambda_D ]}[\nu]) .
\end{eqnarray}

\section{Explicit  calculation}

A full calculation of iterates is usually beyond reach. We can however
get explicit sequences of degrees by considering the images of a
generic projective line in $P_q$, as was introduced in~\cite{FaVi93},
making a link with the geometrical picture of~\cite{Ar90}:

Suppose we start from a restricted diagonal $\Delta_{[\lambda_1,
\lambda_2]}^{(N)}$. It contains $q=N(l_1+l_2)+1$ vertices ${V}_1,
\dots,{V}_q$. For each of these $q$ vertices, we assign to $y$ an
initial value of the form:
\begin{eqnarray}
y_{[{V}_k]} = {{\alpha_k + \beta_k \; x }\over{ \alpha_0 + \beta_0 \; x
}}, \quad  \; k = 1 \dots q
\end{eqnarray}
where $\alpha_0$, $ \beta_0$ and $ \alpha_k$, $\beta_k, (k=1..q)$ are
arbitrary constants, and $x$ is some unknown.  We then calculate the
values of $y$ at the vertices which are within the range of
$\Delta_{[\lambda_1, \lambda_2]}^{(N)}$.  These values are rational
fractions of $x$, whose numerator and denominator are of the same
degree, and that is the degree we are looking for.

The next step is then to evaluate the growth of degrees.  One very
fruitful method is to introduce the generating function of the
sequence of degrees
\begin{eqnarray}
g(s) = \sum_{k=0}^{\infty} s^k \, d^{(k)} 
\end{eqnarray}
and try to fit it with a rational fraction.

{\em The remarkable fact is that it again works surprisingly well, as
it did for maps}, although we know that it may not always be the
case~\cite{HaPr05}. This means that we can often extract the asymptotic
behaviour measured by (\ref{algent}) and (\ref{algent2}) just by
looking at a finite part of the sequence of degrees.

The existence of a rational generating function with integer
coefficients for the sequence of degrees implies that it verifies a
finite recurrence relation. For maps this can sometimes be proved
through a singularity analysis. A similar singularity analysis should
be done here, but it is beyond the scope of this letter.

We have explored a number of examples.  We found examples with non zero
entropy, and examples with vanishing entropy either with linear growth
either with quadratic growth, up to now.

\section{Example 1: the  deformed cross-ratio relation}

Take as a  defining relation:
\begin{eqnarray}
\label{dcr}
\hskip -1cm
f_{dcr} =  {\frac { \left( y_{[{n_1,n_2}]}-a \,y_{[{n_1+1,n_2}]} \right) \left(
 y_{[{n_1,n_2+1}]}-b\,y_{[{n_1+1,n_2+1}]} \right) }{ \left(
 y_{[{n_1,n_2}]}- c\,y_{[{n_1,n_2+1}]} \right) \left(
 y_{[{n_1+1,n_2}]}-d\,y_{[{n_1+1,n_2+1}]} \right) }}-s = 0
\end{eqnarray}
This relation is based on a deformed version of the cross ratio of the
four corner values. It is known to define an integrable lattice
equation for $a=b=c=d$~\cite{NiCa95}.
\noindent

At generic values of the parameters we get the following explicit
values  for $d^{(n)}_{\pm \pm}$ (the defining relation being
very symmetric, the four sequences are the identical):

\begin{eqnarray} 
\{ d^{(n)}_{\pm\pm}\} =  \{1, 2, 4, 9, 21, 50,120, 289, \dots\}.
\end{eqnarray}
This sequence is fitted by the generating function 
\begin{eqnarray}
g_{\pm\pm}^{dcr}(s) = {\frac {1-s-{s}^{2}}{\left(1-s\right) \left(
1-2\,s-{s}^{2} \right)}}.
\end{eqnarray}
The entropy is the logarithm of the inverse of the modulus of the
smallest pole of $g(s)$
\begin{eqnarray}
\epsilon_{\pm\pm}^{dcr} = \log(1+\sqrt {2}).
\end{eqnarray}
We have calculated a number of subsidiary entropies for various values
of $[\lambda_1, \lambda_2]$. They all give sequences which can be
fitted with rational generating functions. The entropies depend on the
``slope'' $\sigma = \lambda_2/ \lambda_1$: large $\sigma$ give larger
$\epsilon_{[\lambda_1, \lambda_2]}[1]$ and smaller
$\epsilon_{[\lambda_1, \lambda_2]}[2]$, and conversely. As an example,
we see that $\epsilon_{[p,1]}[1]$ is the inverse of the logarithm of
the smallest modulus of the roots of $1-s - {s}^{p}-{s}^{p+1}$.  There
is an interesting interplay between the various fundamental and
subsidiary entropies.

\bigskip
For the known integrable case (parameters $a=b=c=d$)~\cite{NiCa95}, we
find the same sequence for all four $\Delta_{\pm \pm}$:
\begin{eqnarray}
\{ d^{(n)}_{\pm\pm}\} = \{1, 2, 4, 7, 11, 16, 22, 29, 37, 46, 56, \dots \},
\end{eqnarray}
fitted by
\begin{eqnarray}
g_{\pm\pm}^{int}(s)& = &{\frac {1-s+{s}^{2}}{ \left( 1-s \right) ^{3}}}
\end{eqnarray}
The growth of the degree is quadratic
\begin{eqnarray}
 d^{(n)}_{\pm\pm} = 1+ \,n \left( n+1 \right)/2 
\end{eqnarray}
and the entropy vanishes.

The few subsidiary entropies we have calculated when $a=b=c=d$ also
vanish, and the degree growth is quadratic.

\section{Example 2: $Q_4$}

We have analysed the so-called $Q_4$ lattice
equation~\cite{Ni02,AdBoSu03,AdSu04}.  The defining relation is given
by:
\begin{eqnarray} 
 && \hskip -2cm A \left( \left( y_{{[n_1,n_2]}}-b \right) \left(
 y_{{[n_1,n_2+1]}}-b \right) - d \right) \left( \left( y_{{[n_1+1,n_2]}}-b
 \right) \left( y_{{[n_1+1,n_2+1]}}-b \right) - d \right) \label{defq4} \\
 && \hskip -2.5cm + B \left( \left( y_{ {[n_1,n_2]}}-a \right) \left(
 y_{{[n_1+1,n_2]}}-a \right) - e \right) \left( \left( y_{{[n_1,n_2+1]}}-a
 \right) \left( y_{{[n_1+1,n_2+1]}}-a \right) - e \right) = f 
 \nonumber
\end{eqnarray}
with
\begin{eqnarray}
\label{q4constr}
\begin{array}{l} 
 d =  (a-b)\;(c-b)  \cr
 e =  (b-a)\; (c -a)    \cr
 f =  A\; B\; C \left( a-b \right)   \cr
 A\; ( c-b ) + B\; ( c-a ) = C\; (a-b) 
\end{array} 
\end{eqnarray}
We have used our approach for generic values of the parameters, that
is to say {\em without}~(\ref{q4constr}). We find for
the fundamental evolutions:
\begin{eqnarray} 
\{ d^{(n)}_{\pm\pm}\} = \{ 1, 3, 7, 13, 21, 31, 43, 57, 73, 91,
111, \dots\}
\end{eqnarray}
 fitted with the generating function
\begin{eqnarray}
g_{\pm\pm}(s) = {\frac {1+{s}^{2}}{\left(1-s\right)^3 }}.
\end{eqnarray}
The growth of the degree is quadratic,
\begin{eqnarray} 
d^{(n)}_{\pm\pm} = 1 + n ( n-1)
\end{eqnarray}
This indicates  integrability of the  form (\ref{defq4}) !

It is interesting to calculate more of the entropies, related to
initial conditions with a different slope, still for unconstrained
parameters. For example
\begin{eqnarray}
&& \hskip -1cm \{ d^{(n)}_{[1,2]}[1] \} = \{ 1, 5, 13, 25, 41, 61, 85,
113, \dots \} \\  && \hskip -1cm \{ d^{(n)}_{[1,2]}[2] \} = \{1, 3,
5, 9, 13, 19, 25, 33, 41, 51, 61, 73, 85, 99, 113,\dots \}
\end{eqnarray}
Both give zero entropy and quadratic growth, as for the fundamental values.

\section{Example 3: Discrete Sine-Gordon}

A multilinear defining relation for the discrete Sine-Gordon equation
can be found in~\cite{Or80,QuCaPaNi91}:
\begin{eqnarray}
\hskip -2cm
y_{[n_1,n_2]} \; y_{[n_1+1,n_2]} \; y_{[n_1,n_2+1]} \;
y_{[n_1+1,n_2+1]}\nonumber \\ - a \; ( y_{[n_1,n_2]} \;
y_{[n_1+1,n_2+1]} - y_{[n_1+1,n_2]} \; y_{[n_1,n_2+1]} ) - 1 =0
\end{eqnarray}
For this lattice equation, we find
\begin{eqnarray} 
\hskip -0cm
\{ d^{(n)}_{\pm\pm}\} = \{ 1, 3, 7, 13, 21, 31, 43, 57, 73, 91,
111, \dots\}\end{eqnarray} as in the previous case (the same quadratic
growth, vanishing $\epsilon_{\pm\pm}$).

The subsequent calculation of $\epsilon_{[1,2]}[1]$ and
$\epsilon_{[1,2]}[2]$ yields 
\begin{eqnarray}
&& \hskip -0cm \{ d^{(n)}_{[1,2]}[1] \} = \{ 1, 4, 11, 21, 34, 51,
71,94,121, 151 \dots \} \\ && \hskip -0cm \{ d^{(n)}_{[1,2]}[2] \} =
\{ 1, 3, 4, 8, 11, 16, 21, 28, 34, 43, 51, 61, 71\dots \}
\end{eqnarray}
fitted respectively by 
\begin{eqnarray}
g_{[1,2]}^{sg}[1] & = & {\frac {1+2\,s+4\,{s}^{2}+2\,{s}^{3}+{s}^{4}}{
 \left( {s}^{2}+s+1 \right) \left(1- s \right) ^{3}}} \\
 g_{[1,2]}^{sg}[2] & = & {\frac {1+2\,s +{s}^{3}+{s}^{5}}{ \left( s+1
 \right) \left( {s}^{2}+ s+1 \right) \left(1- s \right) ^{3}}}
\end{eqnarray}
Which mean vanishing entropies and quadratic growth.

\section{Example 4: Non-isotropic model}
\label{anisotropic}
There are cases where the various directions of evolution are not
equivalent.  The entropies $\epsilon_{\pm\pm} $ are not all equal.
Take the simple  defining relation (see also~\cite{Hi06,HiVi06}):
\begin{eqnarray}
\label{asym}
\hskip -1cm
y_{[ n_1, n_2+1]}y_{[n_1, n_2]}y_{[n_1+1, n_2]}+y_{[ n_1,
n_2+1]}y_{[n_1+1,n_2+1]}+y_{[n_1+ 1, n_2]}= 0
\label{asym}
\end{eqnarray}
The sequences of degrees for the fundamental evolutions differ:
\begin{eqnarray}
\{ d^{(n)}_{[-+]}\} & = & \{1, 3, 7, 17, 41, 99, 239,  \dots\} \\ 
\{ d^{(n)}_{[++]}\} & = & \{1, 2, 4, 7, 14, 28, 56, \dots\} \\ 
\{ d^{(n)}_{[+-]}\} & = &\{1, 2, 5, 10, 20, 40, 80, \dots\} \\ 
\{ d^{(n)}_{[--]}\} & = & \{1, 2, 4, 8, 16, 32, 64, \dots\}
\end{eqnarray}
They   fit with the generating functions
\begin{eqnarray}
g_{[-+]}  =   {{1+s}\over{ 1-2\;s -s^2}} ,& \qquad
& g_{[++]} =    {{ (1-s)( 1+s+s^2) } \over{1-2s}}, \\
g_{[+-]}  =    {{1+s^2} \over{1-2\;s}},& \qquad &
g_{[--]} =  {{1}\over{1-2\;s}},
\end{eqnarray}
so that 
\begin{eqnarray}
\epsilon_{++} = \epsilon_{+-} = \epsilon_{--} = \log(2)
\end{eqnarray}
but
\begin{eqnarray}
 \epsilon_{-+} = \log(2.414...)
\end{eqnarray} 
It does not seem excluded a priori to have vanishing entropy in some
direction and non-vanishing entropy in some other direction, but we
have not exhibited any explicit example of that yet.

\section{Conclusion and perspectives}

The definitions presented here extend to all dimensions ($D > 2$), and
apply to non-autonomous equations, as well as multicomponent systems,
provided the evolutions are rational.
\smallskip

We may use a defining relation which is not multilinear if we do not
insist on having all of  the $2^D$ evolutions described in section
({{5}}). For $D=2$ we may for example accept a defining
relation which is of higher degree in $y_{[n_1+1,n_2]}$ and
$y_{[n_1,n_2+1]}$. The price to pay is to consider only initial
conditions with $\lambda_1 \lambda_2 < 0$.
\smallskip

We may also consider defining relations extending over more than one
elementary cell: this means considering equations of higher
order~\cite{NiPaCaQu92}.
\smallskip

We have done a number of explicit calculations of the fundamental (and
subsidiary) entropies, beyond the ones presented here.  All the
lattice equations which are known to be integrable have vanishing
entropies. These results, and what is already known for maps, suggest
that the vanishing of entropy, that is to say the drastic drop of the
complexity of the evolution, is indeed always a sign of integrability.
\smallskip

Of course the drop of the degree is related to the singularity content
of the evolution, as it was for maps. A systematic
singularity/factorization analysis is one of the tracks to follow. In
this spirit, one can undertake a description of all lattice equations
of a given degree, and for a given dimension. This is the purpose
of~\cite{Hi06,HiVi06}.
\smallskip

One should also determine which of the properties are canonical, that
is to say independent of any change of coordinates one may perform on
the variables. For maps this was just invariance by birational changes
of coordinates on a finite dimensional projective space. Here the
situation is made much more intricate by the infinite number of
dimensions of the space of initial conditions.
\smallskip

All the entropies we have calculated explicitly are the logarithm of
an algebraic integer. There is a conjecture that this is always the
case for maps~\cite{BeVi99,HaPr05}, and we are lead to the same
conjecture here.
\smallskip

Another line of research touches upon arithmetic: a link has been
established between the algebraic entropy and the growth of the height
of iterates, when the parameters and the values of $y_{[n_1, \dots,
n_D]}$ are rational numbers~\cite{AbAnBoHaMa99b,Ha05,AnMaVi05}. This
applies here as well, and will be the subject of further analysis.

\bigskip
\par \noindent {\bf Acknowledgment}: I would like to thank M.~Talon for
constructive discussions.
\bigskip


\begin{thebibliography}{10}

\bibitem{FaVi93}
G.~Falqui and C.-M. Viallet, {\em Singularity, complexity, and
  quasi--integrability of rational mappings}.
\newblock Comm.\ Math.\ Phys. {\bf 154} (1993), pp. 111--125.
\newblock hep-th/9212105.

\bibitem{BeVi99}
M.~Bellon and C-M. Viallet, {\em Algebraic Entropy}.
\newblock Comm. Math. Phys. {\bf 204} (1999), pp. 425--437.
\newblock chao-dyn/9805006.

\bibitem{Ar90}
V.I. Arnold, {\em Dynamics of complexity of intersections}.
\newblock Bol. Soc. Bras. Mat. {\bf 21} (1990), pp. 1--10.

\bibitem{Ve92}
A.P. Veselov, {\em Growth and Integrability in the Dynamics of Mappings}.
\newblock Comm.\ Math.\ Phys. {\bf 145} (1992), pp. 181--193.

\bibitem{HiVi98}
J.~Hietarinta and C.-M. Viallet, {\em Singularity confinement and chaos in
  discrete systems}.
\newblock Phys. Rev. Lett. {\bf 81}(2) (1998), pp. 325--328.
\newblock solv-int/9711014.

\bibitem{TrGrRa01}
S.~Tremblay, B.~Grammaticos, and A.~Ramani, {\em Integrable lattice equations
  and their growth properties}.
\newblock Phys. Lett. {\bf A 278} (2001), pp. 319--324.

\bibitem{Hi81}
R.~Hirota, {\em Discrete Analogue of a Generalized Toda Equation}.
\newblock J. Phys. Soc. Japan {\bf 50} (1981), pp. 3781--3791.

\bibitem{Mi82}
T.~Miwa, {\em On Hirota's difference equations}.
\newblock Proc. Japan Acad. Ser. A Math. Sci. {\bf 58}(1) (1982), pp. 9--12.

\bibitem{QuCaPaNi91}
G.R.W. Quispel, H.W. Capel, V.G. Papageorgiou, and F.W. Nijhoff, {\em
  Integrable mappings derived from soliton equations}.
\newblock Physica {\bf A 173} (1991), pp. 243--266.

\bibitem{HaPr05}
B.~Hasselblatt and J.~Propp.
\newblock {\em Monomial maps and algebraic entropy}.
\newblock arXiv:math.DS/0604521, to be submitted to Ergodic Theory and
  Dynamical Systems,  (2006).

\bibitem{NiCa95}
FW~Nijhoff and HW~Capel, {\em The discrete Korteweg-de Vries equation}.
\newblock Acta Appl. Math. {\bf 39}(1-3) (1995), pp. 133--158.

\bibitem{Ni02}
F.~Nijhoff, {\em Lax pair for the Adler (lattice Krichever-Novikov) system}.
\newblock Phys. Lett. {\bf A 297} (2002), pp. 49--58.
\newblock arXiv:nlin.SI/0110027.

\bibitem{AdBoSu03}
V.E. Adler, A.I. Bobenko, and Yu.B. Suris, {\em Classification of integrable
  equations on quad-graphs. The consistency approach}.
\newblock Comm. Math. Phys. {\bf 233}(3) (2003), pp. 513--543.
\newblock arXiv:nlin.SI/0202024.

\bibitem{AdSu04}
V.E. Adler and Yu.B. Suris, {\em $Q_4$: integrable master equation related to
  an elliptic curve}.
\newblock International Mathematics Research Notice {\bf 2004}(47) (2004), pp.
  2523--2553.
\newblock arXiv:nlin.SI/0309030.

\bibitem{Or80}
S.J. Orfanidis, {\em Group-theoretical aspects of the discrete sine-Gordon
  equation}.
\newblock Phys. Rev. D {\bf 21}(6)Mar 1980, pp. 1507--1512.

\bibitem{Hi06}
J.~Hietarinta.
\newblock {\em Search for factorized lattice maps, Part 1: Results with minimal
  factorization}.
\newblock in preparation.

\bibitem{HiVi06}
J.~Hietarinta and C-M. Viallet.
\newblock {\em Search for factorized lattice maps, Part 2: Algebraic entropy
  analysis}.
\newblock in preparation.

\bibitem{NiPaCaQu92}
F.W. Nijhoff, V.G. Papageorgiou, H.W. Capel, and G.R.W. Quispel, {\em The
  lattice Gel'fand-Dikii hierarchy}.
\newblock Inverse Problems {\bf 8} (1992), pp. 597--621.

\bibitem{AbAnBoHaMa99b}
N.~Abarenkova, J-C.~Angl\`es d'Auriac, S.~Boukraa, and J-M. Maillard, {\em
  Growth-complexity spectrum of some discrete dynamical systems}.
\newblock Physica {\bf D}(130) (1999), pp. 27--42.

\bibitem{Ha05}
R.~Halburd, {\em Diophantine integrability}.
\newblock J. Phys. A {\bf 38}(16) (2005), pp. L263--L269.
\newblock arXiv:nlin.SI/0504027.

\bibitem{AnMaVi05}
J-C.~Angl\`es d'Auriac, J-M. Maillard, and C-M Viallet, {\em On the complexity
  of some birational transformations}.
\newblock J.Phys. A {\bf 39} (2006), pp. 3641--3654.
\newblock arXiv:math-ph/0503074.

\end{thebibliography}

\end{document}